\documentclass[journal, a4paper]{IEEEtran}
\usepackage{amsmath,amsfonts}
\usepackage{algorithmic}
\usepackage{algorithm}
\usepackage{array}
\usepackage[caption=false,font=normalsize,labelfont=sf,textfont=sf]{subfig}
\usepackage{textcomp}
\usepackage{stfloats}
\usepackage{url}
\usepackage{verbatim}
\usepackage{graphicx}
\usepackage{cite}
\usepackage{bm}

\allowdisplaybreaks[4]

\usepackage{tikz}
\newcommand*{\circled}[1]{\lower.7ex\hbox{\tikz\draw (0pt, 0pt)%
    circle (.5em) node {\makebox[1em][c]{\small #1}};}}

\begin{document}

\title{Computing Offloading and Semantic Compression for Intelligent Computing Tasks in MEC Systems}

\author{\IEEEauthorblockN{Yuanpeng Zheng\IEEEauthorrefmark{1}, Tiankui Zhang\IEEEauthorrefmark{1}, Rong Huang\IEEEauthorrefmark{2} and Yapeng Wang\IEEEauthorrefmark{3}
\vspace{0.3cm}
\\\IEEEauthorrefmark{1}\small 
School of Information and Communication Engineering,
Beijing University of Posts and Telecommunications, Beijing 100876, China 
\\\IEEEauthorrefmark{2}\small China Unicom Research Institute, Beijing, China
\\\IEEEauthorrefmark{3}\small Faculty of Applied Sciences, Macao Polytechnic University, Macao SAR, China
\\ \{zhengyuanpeng, zhangtiankui\}@bupt.edu.cn, huangr27@chinaunicom.cn, yapengwang@mpu.edu.mo.
}
\thanks{ This work is supported by Beijing Natural Science Foundation (No.4222010). 
}
}

\maketitle

\begin{abstract}
This paper investigates the intelligent computing task-oriented computing offloading and semantic compression in mobile edge computing (MEC) systems. 
With the popularity of intelligent applications in various industries, terminals increasingly need to offload intelligent computing tasks with complex demands to MEC servers for computing, which is a great challenge for bandwidth and computing capacity allocation in MEC systems. 
Considering the accuracy requirement of intelligent computing tasks, we formulate an optimization problem of computing offloading and semantic compression.
We jointly optimize the system utility which are represented as computing accuracy and task delay respectively to acquire the optimized system utility. 
To solve the proposed optimization problem, we decompose it into computing capacity allocation subproblem and compression offloading subproblem and obtain solutions through convex optimization and successive convex approximation.
After that, the offloading decisions, computing capacity and compressed ratio are obtained in closed forms.
We design the computing offloading and semantic compression algorithm for intelligent computing tasks in MEC systems then. 
Simulation results represent that our algorithm converges quickly and acquires better performance and resource utilization efficiency through the trend with total number of users and computing capacity compared with benchmarks.
\end{abstract}

\begin{IEEEkeywords}
Intelligent computing task, semantic compression, MEC, successive convex approximation.
\end{IEEEkeywords}

\section{Introduction}
With the rapid development of mobile edge computing (MEC), more and more new applications such as computer vision, natural language processing, semantic communication, etc., are emerging constantly in MEC systems\cite{ref3}.
As the increasing number of intelligent computing tasks, MEC needs to tackle with many problems with specific characteristics \cite{ref1,ref6} which is different from traditional resource allocation problems.
However, few existing works consider the various requirements of those characteristics such as compression and computing accuracy \cite{ref2}. 
Therefore, how to efficiently offload to support the specific demands of intelligent computing tasks is still an unaddressed problem.

Hence, in the context of massive Internet of Things (IoT) devices deployment and limited terminal computing capacity, the computing tasks are increasingly complex and highly coupled with communication, existing works on computing offloading and semantic compression in MEC systems has become specific and multidimensional. 
C. Wang \textit{et al.}\cite{ref4} considered computation offloading and content caching strategies in wireless cellular network with MEC and formulate the total revenue of the network. 
With considering edge users and large data volume, G. Faraci \textit{et al.}\cite{ref5} formulated a power consumption and delay optimization problem in unmanned aerial vehicle (UAV) assisted MEC systems. 
Nevertheless, in actual application scenarios, intelligent computing tasks have complicated characteristics and demands which need to be considered in the computing offloading and semantic compression in MEC systems.

As the increasing trend of artificial intelligence, intelligent computing tasks has brought more requirements to the wireless network especially the MEC field. The researches on intelligent computing tasks are getting more attention. 
B. Gu \textit{et al.}\cite{ref2} investigated the fitting of modelling classification accuracy by verification of large data sets for intelligent computing tasks and find that power law is the best among all models.
Considering the scenario of intelligent computing tasks, H. Xie \textit{et al.}\cite{ref11} proposed a brand new framework of semantic communication where a deep learning based system for text transmission combined with natural language processing and semantic layer communication was constructed.
Apparently, conditions for research of computing offloading and semantic compression considering intelligent computing tasks are mature gradually both on application scenarios and modelling of tasks.

There are still some studies considering the demands and characteristics of intelligent computing tasks with semantic compression in MEC systems.
H. Xie \textit{et al.}\cite{ref1} investigated the deployment of semantic communication system based on edge and IoT devices where MEC servers computing the semantic model and IoT devices collect and transmit data based on semantic task model where semantic compression exists on transmission side. 
Y. Wang \textit{et al.}\cite{ref17} proposed a semantic communication framework for textual data transmission and formulated an optimization problem whose goal is to maximize the total semantic similarity by jointly optimizing the resource allocation policy and determining the partial semantic information to be transmitted.
Obviously, the various demands and characteristics of intelligent computing tasks have brought many changes on computing offloading and semantic compression in MEC systems but researches on modelling intelligent computing tasks and applying it to offloading have not been considered yet according to above works.

Obviously, the key challenges of existing works mainly focus on intelligent computing, i.e., intelligent task processing in MEC systems.
Particularly, the combination of computing offloading and semantic compression in MEC systems considering the demands of intelligent computing tasks is still an unaddressed research area.
Based on above works, we focus on resource allocation when task offloading and semantic compression coexist in MEC system.
The main contributions of this paper are as follows. 
We formulate an optimization problem of computing offloading and semantic compression considering accuracy requirement of intelligent computing tasks in MEC systems. We define the system utility which consists of the system revenue depending on computing accuracy and cost depending on task delay. 
The highly coupled computing offloading and semantic compression problem is decoupled into two subproblems including computing capacity allocation subproblem and compression offloading subproblem which are solved by successive optimization approximation and convex optimization. 
The simulation results verify that our algorithm converges quickly and acquires better performance and resource utilization efficiency through the trend with total number of users and computing capacity compared with benchmarks.

\section{System Model}
In order to solve the resource allocation problem of computing offloading and semantic compression in MEC systems, we consider the fog radio access network (F-RAN) scenario and equip MEC servers on small base stations (SBS) to form the MEC systems. We set the total amount of users is $U$. The set of MEC systems is denoted by $K^S = \{1,...,k,...,K\}$ and it is assumed that SBS $k$ is associated with $U_k$ mobile users. We let $U^S_k = \{1,...,u_k,...,U_k\}$ denote the set of users associating with SBS $k$ where $u_k$ refers to the $u$th user which associates with the $k$th SBS. The set of computing tasks is denoted by $M^S = \{1,...,m,...,M\}$ and we consider two types of computing including local computing and offloading to MEC computing in our systems as shown in Fig. 1. 
In our model, step 1 in Fig.1 can be applied to feature extraction in semantic tasks, i.e., semantic compression. Let bandwidth resource of our system be $B$, computing capacity of local device be $F^L_{u_k}$, computing capacity of the MEC server be $F_k$ and delay limit for computing task $m$ be $\tilde{t}_m$. 

\begin{figure}[!t]
  \centering
  \includegraphics[scale=0.9]{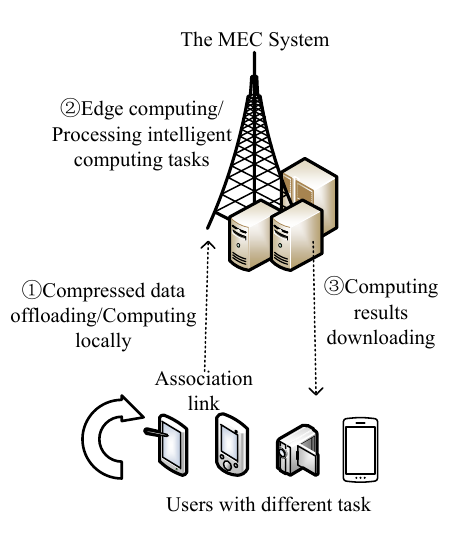}
  \caption{The system model.}
  \label{fig_1}
\end{figure}

\subsection{Communication Model}
In our system, every SBS in the network is equipped with the MEC server, so each user can offload its computing task to the MEC server through the SBS to which it is connected. We denote $x_{u_k} \in \{0,1\}, \forall u,k$ as the computing offloading indicator variable of user $u_k$. Specially, $x_{u_k} = 1$ if user $u_k$ offload its computing task to the MEC server via wireless network and we have $x_{u_k} = 0 $ if user $u_k$ determine to compute its task locally on the mobile device. Therefore, we denote $\bm{x} = \{x_{u_k}\}_{u_k \in U^S_k, k \in K^S}$ as the offloading indicator vector.

In this paper, we consider that spectrum used by SBSs is overlaid and spectrum within one SBS is orthogonally assigned to every user. We only analyze uplink transmission and divide the total spectrum into $N$ subcarriers, which is denoted as $N^S = \{1,...,n,...,N\}$. We denote $\rho_{u_kn}\in \{0,1\},\forall u,k,n$ as subcarrier variables, where $\rho_{u_kn} = 1$ means subcarrier $n$ is allocated to user $u_k$ which is associated with SBS $k$ and $\rho_{u_kn} = 0$ otherwise. Obviously, one subcarrier on an SBS can only be allocated to one user at a time. Then the uplink transmission rate of user $u_k$ on subcarrier $n$  given as
\begin{equation}
  \label{deqn_comm} 
  r_{u_kn} = \frac{B}{N}log_2\left( 1 + \frac{P_{u_kn}g_{u_kn}}{I_{u_kn} + \sigma^2} \right), \forall u,k,n,
\end{equation}
where $P_{u_kn}$ represents transmit power from user $u_k$ to SBS $k$, $g_{u_kn}$ represents wireless channel gain between user $u_k$ and SBS $k$ on subcarrier $n$, and $I_{u_kn} = \sum_{c\in K^S,c\neq k} \sum^{U_c}_{u'_c=1} \rho_{u'_cn}g_{u'_cn}P_{u'_cn},\forall u,k,n$ represents co-channel interference of users associating with other SBSs on the same frequency of user $u_k$. $\sigma^2$ denotes the noise power of additive white Gaussian noise. Obviously, the uplink transmission rate of user $u_k$ is denoted as $r_{u_k} = \sum^N_{n=1} \rho_{u_kn}r_{u_kn},\forall u,k$.

\subsection{Computing Model}
For the computing model, we consider each user $u_k$ has a computing task $m$, and denote $z_{u_km} \in \{0,1\}, \forall u,k,m$ as the indicator variable of the computing task $m$ of user $u_k$. Specially, $z_{u_km} = 1$ if the computing task of user $u_k$ is $m$, otherwise $z_{u_km} = 0$. 
In our model, we assume that $z_{u_km}$ is already given as the user request and we have $\sum^M_{m=1} z_{u_km} = 1,\forall u,k$. 
We consider two types of computing approaches, i.e., local computing and task offloading.

\emph{1) Local Computing:} For the local computing approach, the raw data of user $u_k$ is given as $a_{u_k}$, and we can acquire the computing delay through the raw data $a_{u_k}$ directly, which is given as
\begin{equation}
  \label{deqn_lcd}
  T^L_{u_k} = \frac{\sum \limits ^M_{m=1}z_{u_km}F_{u_km}\left(a_{u_k}\right)}{F^L_{u_k}}, \forall u,k,
\end{equation}
where $F_{u_km}(\cdot ) $ represent the computing resource overhead of the data volume, and we consider it as linear relationship, i.e., $F_{u_km}(a_{u_k} ) = \beta a_{u_k} + \gamma$, where $\beta$ and $\gamma$ are linear paraments.

\emph{2) Task Offloading:} For the task offloading approach, user $u_k$ will compress the raw data $a_{u_k}$ to 
\begin{equation}
  \label{deqn_cd}
  b_{u_k} = \frac{a_{u_k}}{\varepsilon _{u_k}},\forall u,k,
\end{equation}
where $\varepsilon _{u_k}$ is denoted as compression ratio of user $u_k$ and $\varepsilon _{u_k}\geq 1, \forall u,k$. We denote $\bm{\varepsilon} = \{ \varepsilon_{u_k}\}_{u_k\in U^S_k, k\in K^S}$ as the compression ration vector. Apparently, we have $\alpha_{u_k} = (1-x_{u_k})a_{u_k} + x_{u_k}b_{u_k}$. Then, the compressed data $b_{u_k}$ is transmitted to SBS $k$ to process and compute, and the transmission delay of the compressed data from user $u_k$ in wireless link is given as
\begin{equation}
  \label{deqn_trd}
  t^{comm}_{u_k} = \frac{b_{u_k}}{r_{u_k}},\forall u,k.
\end{equation}
Let $f^O_{u_k}$ be computing capacity allocated to user $u_k$ from SBS $k$ and $\bm{f^O} = \{f^O_{u_k}\}_{u_k\in U^S_k, k\in K^S}$ be the computing capacity allocation vector, so that the computing delay of user $u_k$ processing its computing task on SBS $k$ is denoted as
\begin{equation}
  \label{deqn_cd2}
  T_{u_km} = \frac{\sum \limits ^M_{m=1}z_{u_km}F_{u_km}\left(b_{u_k}\right)}{f^O_{u_k}},\forall u,k.
\end{equation}
In this paper, we adopt the same computing resource overhead formula for both raw data and compressed data to represent the same task processed in local devices and MEC servers. Therefore, the delay of computing of user $u_k$ on SBS $k$ is denoted as
\begin{equation}
  \label{deqn_doS}
  t^{comp}_{u_k} = \sum^M_{m=1} z_{u_km}T_{u_km}, \forall u,k.
\end{equation}

We notice the fact that the downlink data volume of computing outcome is much smaller than uplink data volume, so we neglect the downlink transmission in this work. 

\subsection{Utility Function}
To introduce the intelligent computing task feature in our model, we adopt the 3-parameters power law fitting formula between the data volume and the computing accuracy from \cite{ref2} which is the widely used accuracy fitting formula of intelligent classification tasks including semantic compression currently \cite{ref30}. For the convenience of subsequent modelling, we adopt a more simplified form and the computing accuracy of user $u_k$ in our model is denoted as 
\begin{equation}
  \label{deqn_fit}
  y(\alpha_{u_k}) = p - q \alpha_{u_k}^{-r},\forall u,k,
\end{equation}
where $\alpha_{u_k} = (1-x_{u_k})a_{u_k} + x_{u_k}b_{u_k} $ represents the data volume needed computing of user $u_k$ and $p,q,r$ are all fitting paraments. In this work, the limit of computing accuracy of computing task $m$ is set as $\tilde{y}_m$. The total task delay of user $u_k$ in our model is
\begin{equation}
  \label{deqn_tod}
  t_{u_k} = (1-x_{u_k})T^L_{u_k} + x_{u_k}(t^{comm}_{u_k}+t^{comp}_{u_k}), \forall u,k.
\end{equation}

In this paper, we focus on maximum the system utility under computing accuracy constraint and task delay constraint. For each user $u_k$, we consider marginal utility of the combination of system revenue, i.e., computing accuracy and system cost, i.e., task delay. 
We model system utility in the form of a logarithmic function of diminishing marginal utility with the tradeoff of system revenue and cost, therefore the system utility is given as
\begin{equation}
  \label{deqn_su}
  R = \sum_{k\in K^S}\sum_{u_k\in U^S_k} ln\left( L \frac{y(\alpha_{u_k})}{t_{u_k}} \right), \forall u,k,
\end{equation}
where $L$ is denoted as the weight parameter between system revenue and cost. 

\section{Problem Formulation and Algorithm Design}
In order to maximize the system utility, we formulate it as an optimization problem and decompose it into several convex optimization subproblems via successive convex approximation (SCA). 
Then we design the corresponding iterative algorithm to solve the optimization problem.

\subsection{Problem Formulation and Decomposition Solution}
We adopt the system utility proposed in \eqref{deqn_su} as the objective function of our optimization problem, and we formulate it as

\begin{equation}
  \begin{aligned}
  \label{opt}
  &\ \ \, \max_{\bm{x},\bm{f^O},\bm{\varepsilon}} R \\
  &{\rm{s.t.}}\ (\mathrm{C}1): x_{u_k}\in \{0,1\},  \forall u,k, \\
  &\ \ \ \ \ (\mathrm{C}2): \sum^K_{k=1} x_{u_k} \leq 1, \forall u,\\
  &\ \ \ \ \ (\mathrm{C}3): \varepsilon_{u_k} \geq 1, \forall u,k,\\
  &\ \ \ \ \ (\mathrm{C}4):t_{u_k} \leq \sum^M_{m=1} z_{u_km}\tilde{t}_m, \forall u,k, \\
  &\ \ \ \ \ (\mathrm{C}5):y(\alpha_{u_k})\geq \sum^M_{m=1} z_{u_km}\tilde{y}_m, \forall u,k, \\
  &\ \ \ \ \ (\mathrm{C}6):\sum^{U_k}_{u_k =1}f^O_{u_k} \leq F_k, \forall u,k.
  \end{aligned}
\end{equation}

In \eqref{opt}, constraint (C1) guarantees that the value of the computing offloading indicator variables is restrict to 0 and 1, 
constraints (C2) and (C3) means one user can only choose one type of computing approach and the compressed data is less than or equal to the raw data, 
constraints (C4) and (C5) are proposed to ensure the limits of the task delay and the computing accuracy are hold,  
constraint (C6) guarantees that the sum of allocated computing capacity is not greater than total computing capacity of MEC servers.

Obviously, \eqref{opt} is a non-linear mixed integer programming and non-convex optimization problem, and such problems are usually considered as NP-hard problems. Therefore, we need to decompose it into several subproblems and make some transformation and simplification to solve it iteratively. 
For convenience of solving \eqref{opt}, we decompose it into two subproblems by the approach of given variables. 

\emph{1) Computing Capacity Allocation Subproblem:} Under given other variables except $\bm{f^O}$, \eqref{opt} is simplified to
\begin{equation}
  \begin{aligned}
    \label{sub_comp}
    &\ \ \ \ \ \max_{\bm{f^O}} \sum_{k\in K^S}\sum_{u_k\in U^S_k}ln(LA^\delta _{u_k}) - ln (A^\beta _{u_k}+t^{comp}_{u_k})\\
    &{\rm{s.t.}} \ (\mathrm{C}4'): A^\beta _{u_k}+t^{comp}_{u_k} \leq \sum^M_{m=1} z_{u_km}\tilde{t}_m,\forall u,k,\\
    &\ \ \ \ \ (\mathrm{C}6),
  \end{aligned}
\end{equation}
where the constant term $A^\delta_{u_k} = p - q ((1-x_{u_k})a_{u_k} + x_{u_k}b_{u_k})^{-r}$, $A^\beta _{u_k} = (1-x_{u_k})T^L_{u_k} + x_{u_k}t^{comm}_{u_k}$, and $t^{comp}_{u_k} = \frac{\sum^M_{m=1}z_{u_km}F_{u_km}(b_{u_k})}{f^O_{u_k}}$ and in this way, (C4) in \eqref{opt} is converted to $(\mathrm{C}4')$ here. Therefore, \eqref{sub_comp} is a convex optimization problem and can be solved directly by convex optimization method.

\emph{2) Compression Offloading Subproblem:} We need to solve computing offloading indicator variable $\bm{x}$ and compression ratio variable $\bm{\varepsilon}$ under given $\bm{f^O}$. For convenience of solving, we adopt binary variable relaxation and relax $\bm{x}$ into real variable as $x_{u_k} \in \{ 0,1 \}$.
The original problem \eqref{opt} is simplified to
\begin{equation}
  \begin{aligned}
    \label{sub_comoff}
    &\ \ \ \ \, \max_{\bm{x},\bm{\varepsilon}} \sum_{k\in K^S}\sum_{u_k\in U^S_k} ln\left( \frac{L y(\alpha_{u_k})}{(1-x_{u_k})B^\delta_{u_k}+\frac{x_{u_k}}{\varepsilon_{u_k}}B^\beta_{u_k}} \right)\\
    &{\rm{s.t.}} \ (\mathrm{C}1), (\mathrm{C}2),(\mathrm{C}3),\\
    &\ \ \ \ \ (\mathrm{C}4''): (1-x_{u_k})B^\delta_{u_k}+\frac{x_{u_k}}{\varepsilon_{u_k}}B^\beta_{u_k} \leq \sum^M_{m=1} z_{u_km}\tilde{t}_m, \forall u,k,\\
    &\ \ \ \ \ (\mathrm{C}5),
  \end{aligned}
\end{equation}
where the constant terms $B^\delta_{u_k} = T^L_{u_k}$ and $B^\beta_{u_k} = \frac{a_{u_k}}{r_{u_k}} + \frac{\sum^M_{m=1}z_{u_km}F_{u_km}(a_{u_k})}{f^O_{u_k}}$, and $y(\alpha_{u_k}) = p - q ((1-x_{u_k})a_{u_k} + \frac{x_{u_k}}{\varepsilon_{u_k}}a_{u_k})^{-r}$ and in this way, (C4) in \eqref{opt} is converted to $(\mathrm{C}4'')$ here.
Normally, $p,q$ and $r$ satisfy that $p>0, q>0$ and $0\leq r \leq 1$. We adopt the method of variable substitution and let $\eta_{u_k} = 1-x_{u_k}+\frac{x_{u_k}}{\varepsilon_{u_k}}$. Obviously, $\eta_{u_k}$ satisfies that $1-x_{u_k}\leq\eta_{u_k}\leq1$ which will be constraint $(\mathrm{C}3')$ of the above problem and we can transform problem \eqref{sub_comoff} into 
\begin{equation}
  \begin{aligned}
    \label{sub_trans_comoff}
    &\max_{\bm{x},\bm{\eta}} \sum_{k\in K^S}\sum_{u_k\in U^S_k} ln\left( \frac{L (p-q*(a_{u_k}\eta_{u_k})^{-r})}{(1-x_{u_k})(B^\delta_{u_k}-B^\beta_{u_k})+B^\beta_{u_k}\eta_{u_k}} \right)\\
    &s.t. \ (\mathrm{C}1),(\mathrm{C}2),\\ 
    &\ \ \ \ \ (\mathrm{C}3'): 1-x_{u_k} \leq \eta_{u_k}\leq 1, \forall u,k,\\ 
    &\ \ \ \ \ (\mathrm{C}4''): (1-x_{u_k})(B^\delta_{u_k}-B^\beta_{u_k})+B^\beta_{u_k}\eta_{u_k} \leq \\
    &\ \ \ \ \ \ \ \ \ \ \ \ \ \ \ \ \quad \quad \quad \quad \quad \quad \quad \quad \sum^M_{m=1} z_{u_km}\tilde{t}_m, \forall u,k,\\
    &\ \ \ \ \ (\mathrm{C}5): p-q*(a_{u_k}\eta_{u_k})^{-r} \geq \sum^M_{m=1} z_{u_km}\tilde{y}_m,\forall u,k.
  \end{aligned}
\end{equation}

Due to non-convexity of \eqref{sub_trans_comoff}, we adopt the method of SCA and let
\begin{equation}
  \label{deqn_v}
  v_{u_k} \geq ln\left( (1-x_{u_k})(C^\delta_{u_k}-B^\beta_{u_k}) + B^\beta_{u_k}\eta_{u_k} \right).
\end{equation}
We perform first order Taylor expansion on the right side at point $(x^j_{u_k},\eta^j_{u_k})$ and convert it to
\begin{equation}
  \begin{aligned}
  \label{deqn_vv}
  v_{u_k} \geq &ln\left( (1-x^j_{u_k})(B^\delta_{u_k}-B^\beta_{u_k}) + B^\beta_{u_k}\eta^j_{u_k} \right)+\\
  &\frac{(B^\beta_{u_k}-B^\delta_{u_k})(x_{u_k}-x^j_{u_k}) + B^\beta_{u_k}(\eta_{u_k} - \eta^j_{u_k})}{(1-x^j_{u_k})(B^\delta_{u_k}-B^\beta_{u_k}) + B^\beta_{u_k}\eta^j_{u_k}},
\end{aligned}
\end{equation}
which will be constraint (C10) of the above problem. Therefore, \eqref{sub_trans_comoff} is converted to
\begin{equation}
  \begin{aligned}
    \label{sub_sca_comoff}
    &\max_{\bm{x},\bm{\eta}} \sum_{k\in K^S}\sum_{u_k\in U^S_k} ln\left( \frac{L(p-q*(a_{u_k}\eta_{u_k})^{-r})}{v_{u_k}} \right)\\
    &s.t. \ (\mathrm{C}1),(\mathrm{C}2), (\mathrm{C}3'),(\mathrm{C}4''),(\mathrm{C}5),\\
    &\ \ \ \ \ (\mathrm{C}7): \eqref{deqn_vv}.
  \end{aligned}
\end{equation}
Then we can use convex optimization method for SCA iteration to solve \eqref{sub_trans_comoff} by using standard CVX tools\cite{ref29}.

\begin{algorithm}[H]
  \caption{Computing Offloading and semantic Compression Algorithm for Intelligent Computing Tasks}\label{alg:alg1}
  \begin{algorithmic}[1] 
   \STATE Set initial $q=0$, computing offloading variable $x^q$, computing capacity allocation $f^q$ and compression ratio $\varepsilon^q$.
   \STATE Set initial value of $x^0$ and $\varepsilon^0$.
   \STATE Set the iteration constraints $\theta > 0$.
   \REPEAT
   \STATE $q = q + 1$.
   \STATE Obtain $f^q$ by solving computing capacity allocation subproblem \eqref{sub_comp} through $x^{q-1}(t)$ and $\varepsilon^{q-1}$ directly by convex optimization.
   \STATE Obtain $x^q$ and $\varepsilon^q$ by solving compression offloading subproblem \eqref{sub_comoff} through $x^q$ and $f^q$:
   \STATE Set initial $j=0$, the SCA iteration constraint $\theta_2 > 0$, $x^j_{u_k}$ and $\eta^j_{u_k}$ according to \eqref{deqn_vv}.
   \REPEAT
   \STATE $j = j + 1$.
   \STATE Obtain $x^j_{u_k}$ and $\eta^j_{u_k}$ by solving convex optimization problem \eqref{sub_sca_comoff}.
   \STATE Obtain the value of \eqref{sub_comoff}, i.e., $N^j_{sub2}$.
   \UNTIL $\left\lvert N^j_{sub2}-N^{j-1}_{sub2}\right\rvert \leq \theta_2$.
   \STATE Obtain the value of \eqref{opt}, i.e., $N^q$ through $x^q$, $f^q$ and $\varepsilon^q$
   \UNTIL $\left\lvert N^q-N^{q-1}\right\rvert \leq \theta$.
  \end{algorithmic}
\end{algorithm}

\subsection{Algorithm Design and Analysis}
As mentioned above, we decompose the original NP-hard problem \eqref{opt} into two subproblems. Then we use the idea of the greedy algorithm to iterate the above solutions of two subproblems and arrive at the suboptimal solution for \eqref{opt}, which is summarized in Algorithm 1.

In Algorithm 1, we adopt alternating iteration of three problems and obtain the solutions in closed forms by convex optimization. According to the greedy algorithm and convex optimization theory, iteration of two subproblems can ensure $\left\lvert N^q-N^{q-1}\right\rvert \leq \theta$, i.e., convergence quickly but only sub-optimality can be guaranteed \cite{ref23}. 
As we show above, the complexity of Algorithm 1 depends on two subproblems. In subproblem 1, since \eqref{sub_comp} is a convex optimization problem, the complexity is $O(U)$. In subproblem 2, \eqref{sub_comoff} need to be converted to \eqref{sub_sca_comoff} through SCA and achieve solution in iteration algorithm, we assume the number of iterations is $L_{sub2}$, therefore the complexity is $O(UL_{sub2})$. We assume the number of total iteration is $L_{it}$, then the overall complexity of Algorithm 1 is $O((U+UL_{sub2})L_{it})$.
In the way, the NP-hard optimization problem \eqref{opt} is decomposed into low-complexity subproblems and iteratively solved.

\section{Simulation Result}
In this section, we first set the simulation paraments and then show our simulation results to evaluate the performance of our proposed algorithm.

We consider system level simulation of uplink transmission in a small cell F-RAN according to the 3GPP normative document of small cell network, i.e., urban micro (UMi) model \cite{ref25}.
In our model, we consider that four SBSs are deployed in a small cell area with a total coverage of $200{\rm{m}} \times 200{\rm{m}}$. The SBSs provide offloading association and resource allocation for users and note that the path loss depend on the link state of LoS and NLoS \cite{ref25}. 
In our system model, we consider computing accuracy where the paraments of them are set according to the most suitable fitting paraments \cite{ref2}. 
Part of simulation paraments are summarized in Tabel II.
\begin{table}[h]
  \renewcommand\arraystretch{1.5}
  \caption{Part of Simulation Parameters}
  \centering
  \begin{tabular}{p{4cm}<{\centering}|p{2.5cm}<{\centering}}
  \hline
  \hline
  {\textbf{Parameter}} & {\textbf{Value}} \\
  \hline
  Bandwidth resource, $W_k$ & 10 MHz \\
  \hline
  Transmit power, $P_{u_kn}$ & 0.1 W \\
  \hline
  The noise power, $\sigma^2$ & -100 dBm \\
  \hline
  The carrier frequency, $F^q$ & 3.5 GHz\\
  \hline
  Computing capacity of the MEC server, $F_k$ & 200 Gigacycle/s \\
  \hline
  Computing capacity of local device, $F^L_{u_k}$ & 1.4 Gigacycle/s \\
  \hline
  The total number of users, $U$ & 30\\
  \hline
  The weight parameter, $L$ & 1\\
  \hline
  The number of iterations, $L_{it}$ & 10\\
  \hline
  Fitting paraments of computing accuracy, $p,q,r$ & 100, 80, 0.6\\
  \hline 
  \hline
  \end{tabular}
\end{table}

According to the computing delay and accuracy requirements of some services of ultra reliable low latency communications \cite{ref27}, we assume there are three task types in our simulation and the requirements are different. The delay and accuracy limits of tasks are shown in Tabel III.
\begin{table}[h]
  \renewcommand\arraystretch{1.5}
  \caption{Task Parameters}
  \centering
  \begin{tabular}{p{1.5cm}<{\centering}|p{1.6cm}<{\centering}|p{2.8cm}<{\centering}}
    \hline
    \hline
    {\textbf{Task type}} & {\textbf{Task delay}} \bm{$\tilde{t}_m$} & {\textbf{Computing accuracy}} \bm{$\tilde{y}_m$}\\
    \hline
    1 & 20 ms & 85\% \\ 
    \hline
    2 & 40 ms & 90\% \\
    \hline
    3 & 60 ms & 95\% \\
    \hline
    \hline
  \end{tabular}
\end{table}

In order to verify the performance of the proposed algorithm, we add the following schemes for comparison:
\begin{itemize}
\item{Average Computing (AC): The scheme is that computing capacity of MEC servers is allocate averagely.}
\item{Without Compression Ratio (WCR)\cite{ref10}: According to the scheme in \cite{ref10}, the compression ratio is not considered and computing offloading is processed directly.}
\end{itemize}

We demonstrate the convergence of all schemes in Fig. 2 and we can see that the convergence of our proposed algorithm is fast in $L_{it}$ iterations and the trend is basically fixed after convergence, which means our algorithm based SCA and iteration have a good stability and the astringency. From the convergence of comparison algorithm, we find that our proposed algorithm can acquire a better value of system utility and better optimization character in our system model considering joint allocation of communication resource and computing capacity. 
\begin{figure}[!t]
  \centering
  \includegraphics[scale = 0.5]{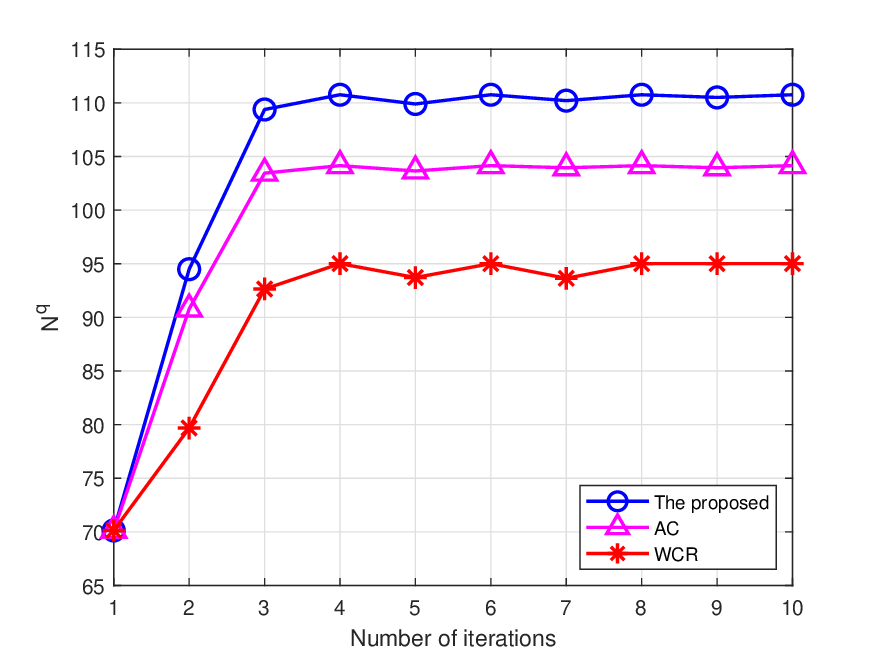}
  \caption{Convergence of all algorithms.}
  \label{fig_2}
\end{figure}

The characteristics of system utility with total number of users $U$ under different bandwidth, i.e., 10 MHz and 50 MHz, as Fig. 3 shows.
It is found that the system utility increases with the total number of users and the trend is slower when total number of users is greater than 35 in our proposed algorithm.
When total number of users is relatively small, the resources are sufficient and resource allocation is efficient, therefore the system utility increases quickly. Nevertheless, as total number of users is relatively large, the resources of system is limited and resource allocation will become inefficient, the growth tendency of system utility will slow down.
The comparison schemes all have this property but the trend is not notable, which is different for different algorithms.
We can see in this figure that the higher bandwidth has larger impact in our proposed algorithm than other comparison schemes which means our scheme have higher usage in bandwidth. 
\begin{figure}[!t]
  \centering
  \includegraphics[scale = 0.5]{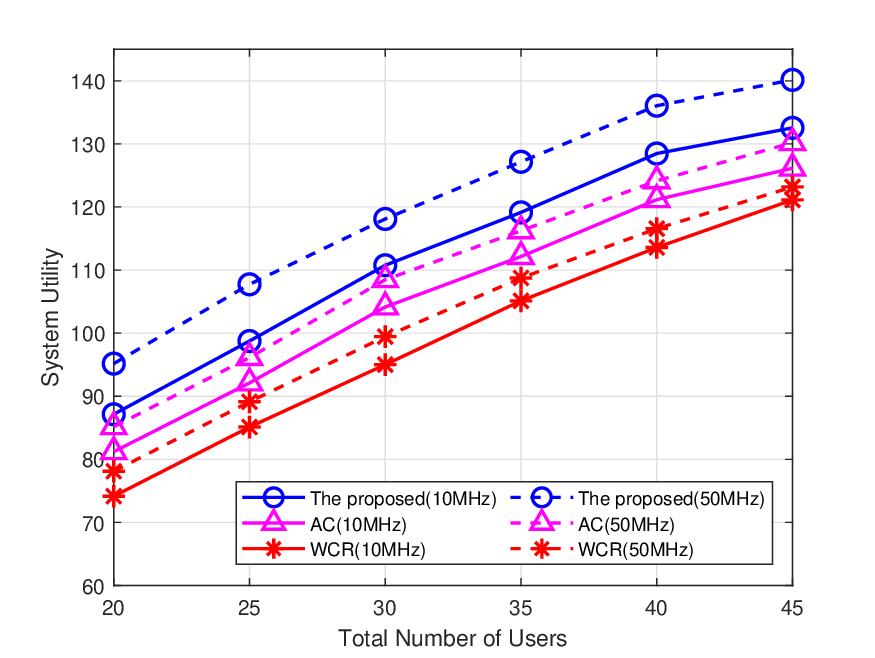}
  \caption{System utility varying with total number of users under different bandwidth.}
  \label{fig_3}
\end{figure}

\begin{figure}[!t]
  \centering
  \includegraphics[scale = 0.5]{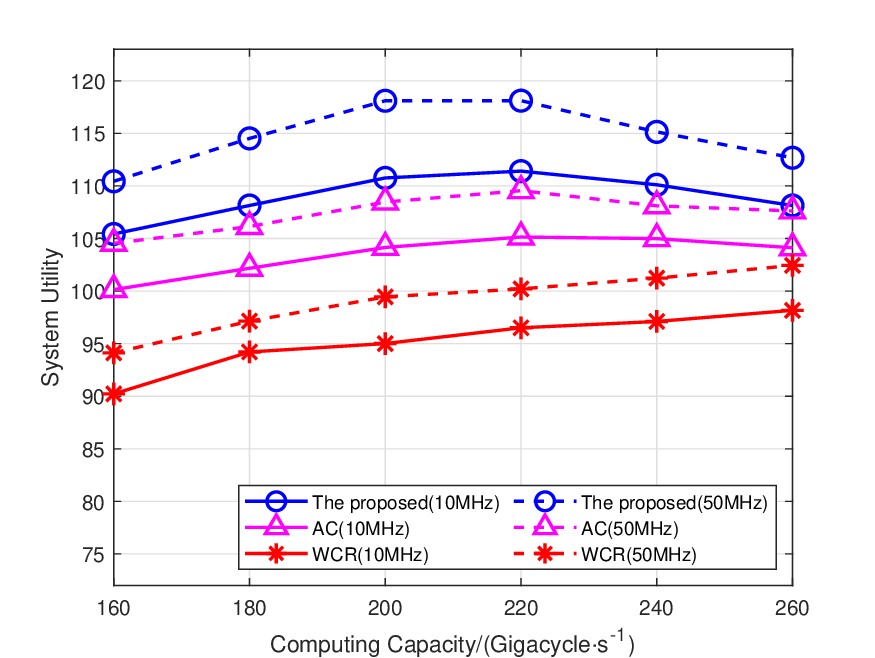}
  \caption{System utility varying with computing capacity of MEC servers under different bandwidth.}
  \label{fig_4}
\end{figure}
We compare system utility with computing capacity $F_k$ of MEC servers under different bandwidth in Fig. 4. From the trend we can see that there is a maximum value of system utility in $F_k = 200$ Gigacycle/s in our proposed algorithm. 
This is because we consider the computing accuracy limit in our system model, the system utility depends on computing accuracy and task delay and our proposed algorithm need make a trade-off between them. We can get a better trade-off when $F_k$ is relatively small an reaches a certain value. However, the communication resource will be limited and affects the compression ration and limits computing accuracy when $F_k$ continues to rise, therefore users would choose local computing which result in the decrease of the system utility.
This property also presents in comparison algorithm AC with the different maximum point, but in WCR where compression ratio is not considered, the trade-off does not existed while $F_k$ is increasing. 
Also, we can see that higher bandwidth do not have a significant impact on this trend of system utility.

Obviously, our methods can be applied to the practical systems that specific intelligent tasks, i.e., semantic compression and computing offloading coexist in MEC systems when there are requirements for computing accuracy and task delay and solve the decision problems of offloading and compression. Our algorithm can obtain higher revenue than traditional methods in this scenario. However, we do not consider communication decision and more general intelligent task computing in the model, resulting in the lack of generality of the application of the model, which will be studied in future works.

\section{Conclusion}
In this paper, we investigated the computing offloading and semantic compression for intelligent computing tasks in MEC systems.
Specially, considering accuracy requirement of intelligent computing tasks, we formulate an optimization problem of computing offloading and semantic compression and decomposed it into two subproblems which were solved iteratively through convex optimization and successive convex approximation.
Simulation results has demonstrated that our algorithm converges quickly and acquires better performance and resource utilization efficiency through the trend with total number of users and computing capacity compared with benchmarks.

\vfill

\end{document}